\documentstyle[12pt]{jarticle}
        \newcommand{\lbc}{\left\langle}
        \renewcommand{\>}{\right\rangle}
        
        \renewcommand{\)}{\right)}
        \renewcommand{\(}{\left(}
        \newcommand{\be}{\begin{equation}}
        \newcommand{\ee}{\end{equation}}
        \newcommand{\bc}{\begin{center}}
        \newcommand{\ec}{\end{center}}
        \newcommand{\bfr}{\begin{flushright}}
        \newcommand{\efr}{\end{flushright}}
        \newcommand{\bfl}{\begin{flushleft}}
        \newcommand{\efl}{\end{flushleft}}

\title{\large \bf INTERMITTENCY AND PHASE TRANSITIONS}
\author{\normalsize E.R. NAKAMURA \\
{\normalsize \it Department of Physics, Fukui Prefectural University,
Matsuoka, Fukui 910-11, Japan } \\
and \\
\normalsize K. KUDO, T. HASHIMOTO and I. YONEDA \\
{\normalsize \it Department of Applied Physics, Fukui University, Fukui 910,
Japan}}
\date{}

\begin{document}

\maketitle
\vspace{2cm}
\begin{quotation}

On the basis of a lattice gas model and the convolution formula with cell
construction scheme, we demonstrate that intermittency in the rapidity-space
with respect to the scaled moments comes from a phase transition between
ordered phase and disordered phase.
It is pointed out that as concerns the power-law behavior of the moments
the critical exponent depends on the size of rapidity interval in the case of
the system which is not enough near to the critical point.
\end{quotation}

\newpage
\begin{center}
{\bf 1. Introduction}
\end{center}

Inspired by the stimulating proposal of intermittency in the rapidity-space
with respect to the factorial moments [1], several authors [2-6] have
presented their theoretical ideas to examine intermittent behavior [7]
from the view point of a phase transition between hadronic matter and
quark-gluon plasma.
Most of the authors investigated this behavior, supposing intermittency of the
moments of rapidity distributions as a striking signal of short-range
fluctuation characterized in cooperative phenomena near the critical point.
However, some authors [5] have considered that such behavior is only to show
the relevance to the short distance correlations.

If intermittent behavior is some signal of critical phenomena, i.e. not a mere
representation of the short distance correlations, this apparent short-range
fluctuation in the rapidity-space is ought to yield the fluctuation of a
macroscopic quantity of state connected with some order parameter.
If so, one should examine that intermittency concerning with the factorial
moments comes from some order parameter which expresses the large fluctuations.
This is a reliable way to ascertain whether one can surely find evidence of
critical behavior in intermittency.

The aim of this paper is to show the universal relation between intermittent
behavior and cooperative phenomena with regard to phase transitions.
We will present a method to obtain the factorial moments of rapidity
distributions from an order parameter in view of the universal correspondence
between the Ising model and the lattice gas model.
In the same manner to illustrate universal singularity of the order parameters
near the critical points of these models, we account the reason why the
factorial moments show intermittency, i.e. a characteristic power-law behavior
in the rapidity-space, on the basis of the convolution formula of the
probability theory [8].
In the framework of the convolution theory it is easy to comprehend that
self-similarity of the cascading models [1,4] and the fractal systems [2,3] is
an extremely restricted condition for the singular factorial moments.
It may be possible to loosen this condition for intermittency in connection
with phase transitions.

In the next section we will present a method to express the factorial moments
of the rapidity distributions in terms of an order parameter, i.e. the
macroscopic fluctuation of particle number in the rapidity interval space,
which is based on the lattice gas model.
This order parameter is equivalent to magnetization in the Ising model.
In sect. 3, according to the convolution theory for phase transitions, we
demonstrate that the power-law behavior of the moments is derived from the
singular probability densities for finding particles.
In sect. 4, we summarize our conclusions.
\begin{center}
{\bf 2. Lattice gas model and intermittency}
\end{center}

It is well known that the lattice gas model for the liquid-gas transition is
mathematically equivalent to the Ising model of ferromagnetic phase transition.
According to the lattice gas model [9], the number of particles of lattice gas
corresponds to the number of down spins in the Ising model.
The $z$ component of the spin $s_i$ at the lattice point $i$ is connected with
the gas density $\rho_i$ as
\be
s_i = 2\( \frac{1}{2} - \rho_i \).
\ee
If the lattice consists of $N$ sites, with the help of eq. (1) magnetization
per lattice site is given by
\be
I = \frac{ \sum_i s_i}{N} = \frac{N_+ - N_-}{N} = \frac{\Delta N}{N},
\ee
where $N_+(N_-)$ is the number of up (down) spins and $\Delta N = N_+ - N_-$
indicates fluctuation of the number of particles.
Putting $N = V$ and $N_- = \sum_i \rho_i$ [9], we obtain from eqs. (1) and (2)
the specific volume per particle
\be
v = \frac{V}{N_-} = \frac{1}{\lbc\rho\>} = \frac{2}{1-I},
\ee
where $\lbc\rho\> = \sum_i \rho_i/N$.
It turns out from above-mentioned correspondence between the lattice gas model
and the Ising model that magnetization can be represented by $I=\Delta N/V$
[10], and hence, if fluctuation of the number of particles becomes large, the
order parameter $I$ ought to be non zero even if volume of the system $V$ is
large enough.

Let us consider the particle number distribution in the rapidity space, and
imagine cell construction of lattice gas according to Kadanoff's construction
of block spins [11].
It is natural to identify $M$ cells of lattice gas with $M$ intervals in the
rapidity space [3].
In the framework of cell construction the total number of particles ($n$)
distributed in $M$ intervals is provided by
\be
\rho_1 + \rho_2 + \cdots + \rho_M = N_- = n.
\ee
The probability for finding particles in the $i$-th interval is $p_i=\rho_i/n$.
In this space we have [1]
\be
p_i=\int_{y_{i-1}}^{y_i} \(\frac{1}{\sigma}\frac{d\sigma}{dy}\)dy\left/
\int_{y_0}^{y_{max}} \(\frac{1}{\sigma}\frac{d\sigma}{dy}\)dy \right.
=H\(y_i\)\left/ M \right.,
\ee
and $M=\Delta y/\delta y =(y_{max}-y_0)/(y_i - y_{i-1})$.
In the same manner as in ref. [1], the scaled moments are defined by
\begin{eqnarray}
\lbc C_l\> & = & \int \cdots \int dp_1 \cdots dp_M P(p_1, \cdots ,
p_M)\frac{1}{M}\sum_{i=1}^M
\(M p_i\)^l \nonumber \\
        & = & \lbc\frac{1}{M} \sum_{i=1}^M \left[ H(y_i)\right]^l\>
          = \lbc\frac{1}{M} \sum_{i=1}^M \( M p_i\)^l\>
\end{eqnarray}
with the distribution of the probabilities $P(p_1, \cdots , p_M)=\sum_{i=1}^M
\delta \(p_i - H(y_i)/M\)$.

In our lattice gas model the probability for finding particles in the $i$-th
interval is given by $p_i = \rho _i /n = 0$ for $\rho_i = 0$, or $p_i =
\rho_i/n=1/n$ for $\rho_i = 1$, where $\rho_i$ means renormalized effective
particle density under the regime of cell construction.
This lattice gas model yields the scaled moments as follows;
\be
\lbc C_l\>=f_l(M)=\lbc\frac{1}{M} \sum_{i=1}^M \( M
p_i\)^l\>=\(\frac{M}{n}\)^{l-1}.
\ee
In order to see the relation between this model and the original Ising model we
set $M=N$ and $n=N_-$.
With the help of eq. (3) it is possible to rewrite eq. (7) into
\be
\lbc C_l\>=f_l(N)=\( \frac{N}{N_-} \)^{l-1}=v^{l-1}.
\ee
Therefore, we get from eqs. (2), (3) and (8) the scaled factorial moments in
terms of an order parameter $I$ in the following form;
\be
\ln \lbc C_l\>=\ln f_l(I)=(l-1)\ln \(\frac{2}{1-I}\),
\ee
which is the increasing function of $I$ $( \mid I\mid \leq 1)$.
\begin{center}
{\bf 3. Convolution formula for intermittency}
\end{center}

Let us consider random separation of rapidity interval $\Delta y_i$ $(i=1,
\cdots, M)$ in $\nu$ segments like
\be
\Delta y_i =y_1 +y_2+\cdots + y_{\nu}.
\ee

Since these segments are independent random variables, the probability
$p_i(\Delta y_i)$ for finding particles in this rapidity interval is given by
convolution of the respective probabilities $p_{ij}(y_j)$ in each segment as
follows [8];
\be
p_i\(\Delta y_i\)=\sum_{\nu} Q(\nu)\int \cdots \int p_{i1}(\Delta y_i - \sum
_{j=2}^{\nu}
y_j) p_{i2}(y_2) \cdots p_{i\nu}(y_{\nu})dy_2\cdots dy_{\nu},
\ee
where $Q(\nu)$ is the weight function to convolute each probability.

As previously mentioned, in our lattice gas model the probability for finding
particles in the rapidity interval is directly connected with the particle
number density, and so the spin variable.
As concerns phase transition between ordered phase and disordered phase, these
variables construct a sequence of them according to cell construction, and
hence approach to the fixed value near the critical point [12].
Thus, they lead to the critical singularity of the observed physical
quantities, i.e. the power-law behavior of them.

If we set the asymptotic value of the probability density near the critical
point as $p_c(y)$, and put $\int p_c dy=M^{-1/\nu}W_c$ with $\Delta y_i =\nu_i
y$ and $ M=(y/\delta y)\sum _{i=1}^M \nu_i$, we obtain from eq. (11)
\be
p_i(\Delta y_i)=\sum_\nu Q(\nu)\frac{1}{M}(W_c)^{\nu}.
\ee
Accordingly, the moments (6) near the critical point are given by
\be
\lbc C_l\>=\lbc \frac{1}{M}\sum_{i=1}^M (Mp_i)^l\>=\sum_{\nu}Q(\nu)(W_c)^{\nu
l}.
\ee
As for independent separation of rapidity intervals, the Poisson distribution
$Q(\nu)=\( \lbc\nu\>^\nu/\nu ! \) \exp (-\lbc\nu\>)$ yields
\be
\lbc C_l\>=\exp \left\{ \lbc\nu\> \left[ (W_c)^l -1 \right]
\right\}=M^{\phi_l},
\ee
where $\phi_l=\lbc\nu\>[(W_c)^l -1]/\ln M$.
Assuming $[(W_c)^l-1] \approx \ln (W_c)^l$ for $(W_c)^l \approx 1$ and
$\lbc\nu\>=\ln M/\ln \lambda$, we have $\phi_l=\ln (W_c)^l/\ln \lambda$.
This is the case of recurrent self similar separation of rapidity intervals in
$\lambda$ segments based on the fractal structure in the rapidity space [1].

The self similar separation model demands $W$ to be constant in this space,
which provides the constant critical exponent $\phi$.
However, it turns out from our model that around the critical point $W$ depends
on the finite lattice size ($y$) in the strict sense, and so $W$ may depend on
rapidity interval ($\delta y$).
Therefore, it is probable that the critical exponent $\phi$ slightly depends on
$\delta y$ through $W$, if the system does not approach enough near to the
critical point.

\newpage
\begin{center}
{\bf 4. Conclusions}
\end{center}

We have shown that intermittency, i.e. fluctuations of the number of particles
in different sizes in the rapidity-space, surely reflects a phase transition
between ordered phase and disordered phase.
We have presented a lattice gas model which easily leads us to comprehend that
fluctuation of the number of particles represents the order parameter like
magnetization of spin systems, and the scaled moments are given only by this
order parameter.
We, thus, have concluded that intermittency with respect to the moments results
from cooperative phenomena of the system near the critical point.

Intermittency of the system is characterized by the power-law behavior of the
scaled moments like the power-law singularity of the order parameters in
various phase transitions.
This behavior of the moments is derived from the critical value of the
probability number density in the rapidity-space on the basis of the
convolution formula with cell construction scheme.
In contrast with other results obtained from usual self-similar fractal models,
we have pointed out that the critical exponent should depend on the lattice
size, i.e. the size of rapidity interval, if the system is not in the condition
enough near to the critical point.
Recent experimental data [13] support this consequence.

\newpage
\begin{center}
{\bf References}
\end{center}
\begin{enumerate}
\item[{[1]}]
A.Bialas and R. Peschanski, Nucl. Phys. B273 (1986) 703; B308 (1988) 857
\item[{[2]}]
J. Wosiek, Acta Phys. Pol. B19 (1988) 863
\item[{[3]}]
H. Satz, Nucl. Phys. B326 (1989) 613;
B. Bambah, J. Fingberg and H. Satz, B332 (1990) 629
\item[{[4]}]
C.B. Chiu and R,C, Hwa, Phys. Lett. B236 (1990) 466
\item[{[5]}]
J. Dias de Deus and J.C. Seixas, Phys. Lett. B246 (1990) 506;
J. Dias de Deus, Phys. Lett. B278 (1992) 377
\item[{[6]}]
N.G. Antoniou, Phys. Lett. B245 (1990) 624
\item[{[7]}]
T.H. Burnett et al. (JACEE), Phys. Rev. Lett. 50 (1983) 2062;
G.J. Alner et al. (UA5), Phys. Lett. B138 (1984) 304;
M. Adamus et al. (NA22), Phys. Lett. B185 (1987) 200;
M.I. Adamovich et al. (EMU-01), Phys. Lett. B201 (1988) 397;
W. Braunschweig et al. (TASSO), Phys. Lett. B231 (1989) 548;
R. Holynski et al. (EMU-07), Phys. Rev. Lett. 62 (1989) 733
\item[{[8]}]
E.R. Nakamura and K. Kudo, Phys. Rev. D41 (1990) 281
\item[{[9]}]
T.D. Lee and C.N. Yang, Phys. Rev. 87 (1952) 410
\item[{[10]}]
M. Suzuki, Prog. Theor. Phys. 69 (1983) 65
\item[{[11]}]
L.P. Kadanoff et al., Rev. Mod. Phys. 39 (1967) 395
\item[{[12]}]
K.G. Wilson and J. Kogut, Phys. Rep. 12C (1974) 75
\item[{[13]}]
M.I. Adamovich et al. (EMU-01), Z. Phys. C-Particles and Fields 49 (1991) 395
\end{enumerate}
\end{document}